\documentclass[aps,prb,twocolumn,
showpacs]{revtex4}
\usepackage{ae}
\usepackage[T1]{fontenc}
\usepackage[ansinew]{inputenc}
\usepackage{amsmath}
\usepackage{amssymb}
\usepackage{graphicx}
\usepackage{color}
\usepackage[colorlinks]{hyperref}
\usepackage{epstopdf}

\begin{document}

\date{\today}



\title{Giant oscillations of the current in a dirty 2D electron
system flowing perpendicular to a lateral barrier under magnetic
field.}

\author{A. M. Kadigrobov}
\affiliation{$^{1}$ Theoretische Physik III, Ruhr-Universit\"at
Bochum, D-44801 Bochum, Germany}

\begin{abstract}
The charge transport in a dirty 2-dimensional  electron system
biased  in the presence of a lateral potential barrier under
magnetic field is theoretically studied. The quantum tunnelling
across the barrier  provides the quantum interference of the edge
states localized on its both sides that results  in giant
oscillations of the charge current flowing perpendicular to the
lateral junction.
 Our theoretical analysis is in a good agreement  with the experimental observations
presented  in Ref. \cite{kang}.  In particular, positions of the
conductance maxima  coincide with the Landau levels while the
conductance itself is essentially suppressed even at the energies
at which the resonant tunnelling occurs and hence these puzzling
observations can be resolved  without taking into account the
electron-electron interaction.
\end{abstract}


 \maketitle
\section{Introduction.}

Investigations of low dimensional electronic structures  have
opened new fields in condense matter physics such as
Berezinskii-Kosterlitz-Thouless phase transition
\cite{Berezinskii,KT}, the quantum Hall effect \cite{QH},
   the macroscopic quantum tunnelling
   \cite{Devoret}, the conductance quantization in
   QPCs \cite{Heinzel,QH,Beenakker}, to name a few.

Energy gaps in  electronic spectra in semiconductors and
insulators play a crucial role in their kinetic  and optic
properties, and one of the fascinating features of low dimensional
structures is a possibility to get energy gaps in electronic
spectra which are gapless in the three dimensional case. One such
example is a two dimensional electron gas (2DEG) with a lateral
barrier under an external magnetic field. In this case the
spectrum of electrons skipping along the barrier is an alternating
series of extremely narrow  bands and gaps
\cite{kang,barrier,graphene}, the widths of which being $\sim
\hbar \omega_H$ for the barrier transparency $D\sim 1$. Such a
drastic change of the spectrum is due to the quantum interference
of the edge states located on the opposite sides of the barrier,
the spectra of  the latter being
 gapless in the absence of the tunnelling. Such a spectra of
 alternating narrow bands and gaps arises  if the
quantum interference of the electron wave functions with
semiclassically large phases takes place. The most prominent and
seminal phenomenon of this type is the magnetic breakdown
 phenomenon\cite{Cohen,KaganovSlutskin,FTT,Slutskin} in which large
semiclassical orbits of electrons under magnetic field are coupled
by quantum tunnelling through very small areas in the momentum
space. Other systems  with analogous quantum interference are
those with multichannel reflection of electrons from sample
boundaries \cite{reflection,physica}, samples with grain
\cite{Peschanski} or twin boundaries \cite{Koshkin}. Common to all
these systems are analogous dispersion equations which are sums of
$2 \pi$ periodic trigonometric functions of semiclassically large
phases of the interfering wave functions.

 Dynamics and kinetics of
electrons in 2DEG in the presence of a lateral barrier under
magnetic field $H$ was experimentally and theoretically
investigated in the situation that the current flows
along\cite{barrier,graphene} and perpendicular
\cite{kang,barrierperp} to the barrier in ballistic samples. In
all these cases giant conductance oscillations have been shown to
arise. In paper \cite{barrier} a \textbf{ballistic} sample with a
lateral barrier in the quantum Hall regime was considered. It has
been analytically shown that the lateral junction placed
perpendicular to the current serves as a unique quantum-mechanical
scatter for propagating magnetic edge states, and that  this
quantum "anti-resonant" scatter provides an essential increase of
the transverse conductance as soon as the Fermi energy is inside
one of the energy gaps in the spectrum of electrons skipping along
the junction.

The object of the present paper is to investigate transport
properties of a \textbf{dirty} biased 2DEG with a lateral barrier
under semiclassical magnetic field, the barrier being placed
perpendicular to the current. In contrast to the ballistic
situation considered in paper \cite{barrierperp}  the contribution
of the conventional edge states  to the conductance  is neglected
assuming the following inequalities being satisfied: $\omega_H
\tau\gg 1; \; L_y \gg l_0 (\omega_H \tau) $ where $\omega_H = e
H/mc$ and $l_0=v_F \tau$ are the cyclotron frequency and  the
 electron free path length, respectively, while $\tau$ is
 the free path time and $v_F$ is the Fermi velocity.
 It is shown that the
above-mentioned bands and gaps manifest themselves by giant
oscillations of the transverse conductance with a change of the
magnetic field of the gate voltage.
Detailed analysis of the phases of the tunnelling matrix
and the gap positions shows that the conductance peaks coincide
with Landau levels in agreement with observations presented  in
paper \cite{kang}.

\section{Formulation and solution of the problem.}

\begin{figure}
  \centerline{\includegraphics[width=8.0cm]{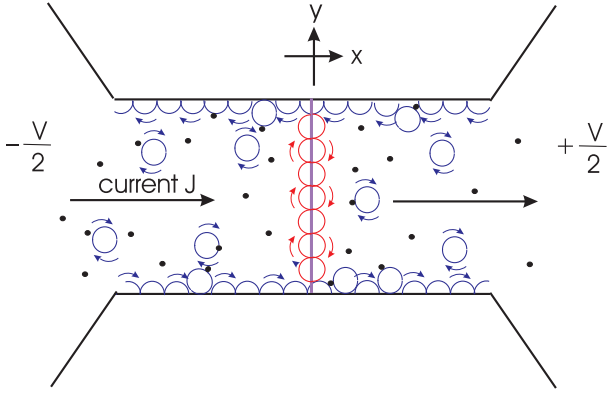}}
  \caption{Schematic presentation of the sample. }
\label{sample1}
  \end{figure}

Let us consider a 2D dimensional electron system in the presence
of a lateral barrier subject to an external  magnetic field $H$
applied  perpendicular to its plane as is shown in
Fig.\ref{sample1}.

In this paper electron dynamics and kinetics are considered in the
semiclassical approximation that is $\hbar \omega_H \ll
\varepsilon_F$ where  $ \omega_H =e H/m c$ is the cyclotron
frequency and $\varepsilon_F$ is the Fermi energy. It is also
assumed that the electron free path length  $l_0 \gg R_h$ where
$R_H = c p_F/eH$ is the Larmour radius and $p_F$ is the Fermi
momentum; the width of the sample $L_y \gg l_0^2/R_H$ and hence
the contribution of the conventional edge states (which are
localized at the external boundaries) to the sample conductance is
negligible. The $x$-axis is parallel to the current flowing along
the biased sample while the $y$-axis is along the lateral barrier.

%
%

\subsection{Dynamics of quasi-particles skipping along the
lateral junction under magnetic field. \label{subsectdynamics}}

As is seen in Fig.\ref{sample1} electrons are in three
qualitatively different states: a) there are  electrons in the
Landau states moving along closed orbits, b) those in the
conventional edge states skipping along the external boundaries of
the sample, and c) electrons in peculiar field-dependent
quasi-particle states highly localized around the lateral barrier.
The quantum interference between the left and right edge states
results in  peculiar one-dimensional spectrum arises.

As shown in Appendix\ref{Appendmatching}, at low transparency of
the barrier $D=|t|^2 \ll 1$ the dispersion equation which
determines the energy $E_n(P_y)$ of quasi-particles skipping along
the barrier is
\begin{eqnarray}
 \cos\bar{\theta}_1(E,P_y)\cos\bar{\theta}_2(E,P_y)=
\frac{|t|^2}{4}\cos\left(\bar{\theta}_{-}(E,P_y)\right)
\label{dispersioneq}
\end{eqnarray}
where $\bar{\theta}_{1,2}=\theta_{1,2} -\pi/4$ and
$\bar{\theta}_{+}=\bar{\theta}_{1}+\bar{\theta}_{2}$ while  $P_y$ is the
conserving projection of the generalized electron momentum on the
direction of the lateral barrier while
\begin{eqnarray}
\theta_1=\frac{c}{e H
\hbar}\int_{-p_E}^{P_y}\sqrt{p_E^2-\bar{p}_y^2 }d
\bar{p}_y;\nonumber \\
\theta_2=\frac{c}{e H
\hbar}\int^{p_E}_{P_y}\sqrt{p_E^2-\bar{p}_y^2 }d \bar{p}_y;\
 \label{theta}
\end{eqnarray}
Here $p_E=\sqrt{2 m E}$. As one easily sees $\bar{\theta}_{+}(E)$
does not depend on $P_y$.

\begin{figure}
  \centerline{\includegraphics[width=8.0cm]{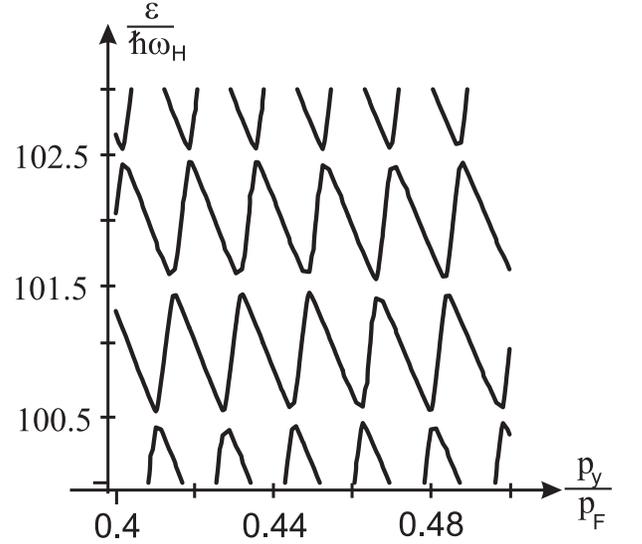}}
  \caption{Spectrum of the quasi-particles skipping along the lateral junction under
  magnetic field. Due to the quantum interference of the edge states
on the left and right sides of the junction this spectrum  is a
series of alternating
  extreme narrow bands  and gaps, the widths of which being  $\sim \sqrt{(1-D)} \hbar
  \omega_H$ and $\sim \sqrt{D}\hbar
  \omega_H$, respectively ($D$ is the barrier transparency). The numerical
  calculations are carried out for $D=0.1$ and the semiclassical parameter
  $\hbar \omega_H/\varepsilon_F =10^{-2}$}
\label{spectrum}
  \end{figure}

One sees from  Eq.(\ref{dispersioneq}) that at $t=0$ there are a
large number of crossing points of the spectra of the left and
right independent edge states (see Appendix \ref{Classification}).
At final barrier transparency   the degeneracy is lifted that
opens gaps $\Delta \sim |t| \hbar \omega$ in the quasi-particle
spectrum (see Fig.\ref{spectrum}).
%
%

As follows from Eq.(\ref{dispersioneq})   degeneration takes place
at $|t|=0$ if two equations are satisfied:
\begin{eqnarray}
\cos\bar{\theta}_1(E,P_y)=\cos\bar{\theta}_2(E,P_y)=0
 \label{degenerpoints}
\end{eqnarray}
Hence   positions of the degeneration points in the $(E,P_y)$
plane are determined by the conditions:
$\bar{\theta}_1(E)=\frac{\pi}{2}(2 k +1)$ and
$\bar{\theta}_2(E)=\frac{\pi}{2}(2 l +1)$ where $k,l $ are
integer. Summing and subtracting them with the use of
Eq.(\ref{theta}) one gets
\begin{eqnarray}
\Big \{    \begin{matrix}
\theta_{+}=E_n/\hbar \omega_H =\pi (n+\frac{1}{2});  \\
\theta_-(E_n,P_y)=\pi(2 k -n+1);
\end{matrix}
 \label{DegPoints}
\end{eqnarray}
where $\theta_{\pm}=\theta_2(E,P_y) \pm \theta_1(E,P_y$ and $n, k
=0,1, ...$.

Therefore, the degeneration points  are in line with  discrete
Landau levels $E_n$  being situated in discrete points $P_y =P_k$
inside the each Landau level $n$ and hence their positions may be
uniquely classified with two discrete indexes  as $P_y=P_k^{(n)}$.

 One easily sees
that the distances between neighboring points are
\begin{eqnarray}
\delta P^{(n)}_k =|P^{(n)}_k-P^{(n)}_{k\pm 1}|\sim
\frac{\hbar}{R_H}\ll p_F
 \label{DegPointsDifference}
\end{eqnarray}

Therefore, if   $Q(P^{(n)}_k , E_n)$   is a slow varying function
of the momentum  on the $\hbar/R_H$ scale one may changes the
summation  with respect to $k$ to the integration as follows:
\begin{eqnarray}
\sum_k Q(E_n, P^{(n)}_k ) = \nonumber \\ - \frac{c}{\pi e \hbar
H}\int Q(E_n, P_y)\sqrt{2 m E_n -P_y^2}dP_y
 \label{variablechange}
\end{eqnarray}

Summing up, the spectrum of electrons skipping along the junction
is an alternating sequence of narrow energy bands $E_n (P_y)$ (the
widths of which are $\sim \sqrt{1-|t|^2}\hbar \omega_H$) and
energy gaps $\Delta_n \sim |t|\hbar \omega_H$ (see
Ref.\cite{kang,barrier,graphene}), the latter lining discrete
Landau levels (n is the  Landau number).

%
%
  \begin{figure}
  \centerline{\includegraphics[width=8.0cm]{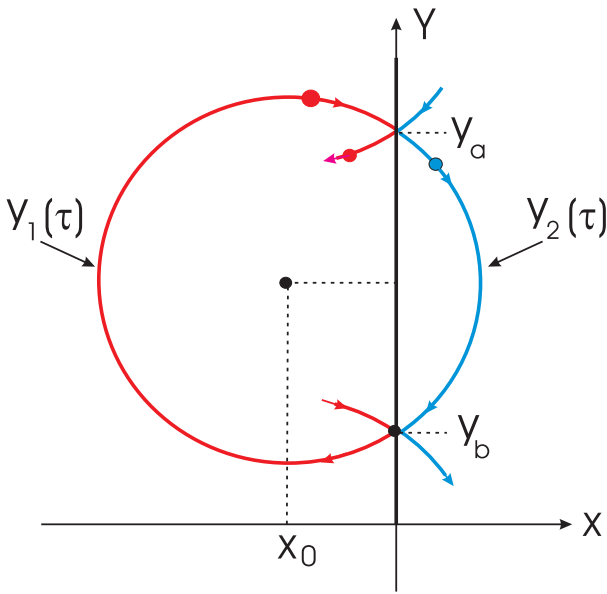}}
  \caption{Schematic presentation of semiclassical motion of a quasi-particle  skipping
  along a lateral barrier (black vertical line) under magnetic field. An electron packet
  moves along a semiclassical orbit,  the $x$-coordinate of its  center
being  $x_0=c P_y/e H$. Due to the quantum-mechanical tunnelling
through the barrier the incoming  packet undergoes a two channel
scattering at the barrier (points $y_{a,b}$) where it is split to
two packets with the amplitude probabilities $t$ and $r$,
$|t|^2+|r|^2=1$. The number of these packets grow in time and
their quantum interference results in peculiar spectrum
Eq.(\ref{dispersioneq}) }\label{orbits}
  \end{figure}

\subsection{Current flowing  along dirty sample and perpendicular to
lateral junction under magnetic field. \label{subsectcurrent}}

As in the case of the magnetic breakdown phenomenon, dynamic and
kinetic properties of quasi-particles skipping along the junction
under magnetic field are of the fundamentally quantum mechanical
nature due to the quantum interference of their wave functions
with semiclassically large phases. Thus, in order to analyze the
transport properties of the  quasi-particles in the presence of
impurities  it is convenient to start with  the equation for the
density matrix $\hat{\rho}$ in the $\tau$-approximation:
%
%
\begin{eqnarray}
\frac{i}{\hbar}\Big[\hat{\rho},\hat{H}\Big]+\frac{\hat{\rho} -
f_0(\hat{H})}{\tau}=-\frac{i
e}{\hbar}\Big[\hat{\rho},V(\hat{x})\Big];
 \label{matrixEq}
\end{eqnarray}
%
%
Here, $\hat{H}$ is the Hamiltonian of the system under
consideration in the absence of the bias voltage $V(x)$, $f_0$ is
the Fermi distribution function,   $\tau $ is the electron
scattering time.

In this paper we assume that the barrier transparency $|t|^2 \ll
1$ is so low that the main drop of the voltage applied to the
sample takes place on the lateral barrier, that is it may be
written as
\begin{eqnarray}
V(x)= V_0 \Theta(-x)
 \label{voltage}
\end{eqnarray}
where $V_0$ is the voltage drop on the barrier and $\Theta$  is
the unit step function.

Writing the density matrix in the form $\hat{\rho}= f_0(\hat{H})+
\hat{\rho}^{(1)}$ and linearizing Eq.(\ref{matrixEq}) with respect
to the bias potential  one gets
%
%
\begin{eqnarray}
\Big[\hat{\rho}^{(1)},\hat{H}\Big]- \frac{i \hbar
}{\tau}\hat{\rho}^{(1)} =-e\Big[f_0(\hat{H}),V(\hat{x})\Big];
 \label{matrixEqLinear}
\end{eqnarray}
%
%

In terms of the density matrix the  the current density at a point
${\bf r}_0$ is written as follows:
%
%
%
%
%
%
\begin{eqnarray}
\textbf{J}=2 e \text{Tr}\left\{\hat{\textbf{v}} \hat{\rho}\right\}
 \label{currenttrace}
\end{eqnarray}
%
%
where $\hat{{\bf v}}$ is the  operator of the quasi-particle
velocity.

Taking matrix elements of  equation Eq.(\ref{matrixEqLinear}) with
respect  to the proper functions of the Hamiltonian $\hat{H}$
written in the Dirac notations
%
%
\begin{eqnarray}
\hat{H}\left|n, P_y\right \rangle=E_n(P_y) \left|n,P_y\right
\rangle;
 \label{DiracNotation}
\end{eqnarray}
%
%
 one finds the density matrix. Inserting the found solution in
Eq.(\ref{currenttrace}) one obtains the current $J$  flowing
perpendicular  to the barrier as follows:
%
%
%
%
%
%
%
\begin{eqnarray}
J=-i 2 e^2\hbar \nu_0 \frac{L_y}{L_x} \sum_{n \neq n^{\prime}}\int
\frac{dP_y}{2\pi \hbar} \;
\frac{V_{n,n^{\prime}}(P_y)v^{(x)}_{n^{\prime},n}(P_y)}{\big[E_{n^{\prime}}
(P_y)-E_n (P_y)\big]^2} \nonumber \\
\times \Big[ f_0\Big(E_n (P_y)\Big) - f_0\Big(E_{n^{\prime}}
(P_y)\Big)\Big];
 \label{currentmatrix}
\end{eqnarray}
%
where $\text{O}_{n,n^{\prime}}(P_y)=<n,
P_y|\hat{\text{O}}|n^{\prime}, P_y>$ and $\nu_0 =1/\tau$ is the
electron-impurity relaxation frequency. The equation is written
under assumption that $\nu_0 \ll |t| \omega_H$. Matrix elements of
the applied voltage and  the velocity operator are presented in
Appendix (see Eqs.(\ref{Vnn},\ref{matrixvelocityfinal})).

Such a peculiar dynamics as is  described in Subsection
\ref{subsectdynamics} manifests itself in  the resonant properties
of the matrix elements in Eq.(\ref{currentmatrix}) that is
especially pronounced at $|t|\ll 1$. Consider, e.g.,
$V_{n,n^\prime}$, Eq.(\ref{Vnn}). At $|t|^2=0$ the wave functions
$\Psi_2$ in Eq.(\ref{Vmatr}) are orthogonal edge state functions
and  hence $V_{n,n^\prime}(P_y)=0$ at $n \neq n^\prime$.
Therefore, one has $V_{n,n^\prime}(P_y)\neq 0$ exclusively due to
a final barrier transparency. Using Eq.(\ref{dispersioneq}) one
easily finds that far from the degenerate points $V_{n,n^\prime}
\propto |t|^2$ while in the vicinity of them the resonance
tunnelling  takes place and $V_{n,n^\prime} \propto 1/2$ and hence
the main contribution to the integral in Eq.(\ref{currentmatrix})
 is from $P_y$ in the vicinity of the degenerate points $P^{(n)}_k$
 (see Eq.(\ref{DegPoints}).

Therefore, when calculating the current Eq.(\ref{currentmatrix})
one may use
Eqs.(\ref{VnnDegeneration},\ref{mevelocitydegeneration}) and get
it as follows:
%
\begin{eqnarray}
J=-2e^2 \nu_0V_0 \frac{L_y}{L_x} \sum_{n \neq n^{\prime}}\int
\frac{dP_y}{2\pi \hbar} T_1(\kappa) C_1^{*}(\kappa) C_2(\kappa^\prime)\nonumber \\
\sum_{\alpha=1}^{2}C_{\alpha}^{*}(\kappa^{\prime})
C_{\alpha}(\kappa)X_\alpha (\kappa)\frac{f_0\left(E_n (P_y)\right)
- f_0\left(E_{n^{\prime}} (P_y)\right)}{E_{n} (P_y)-E_{n^{\prime}}
(P_y)};
 \label{CurrentConstants}
\end{eqnarray}
%
where $\kappa =\{n, P_y\}$ and $\kappa^\prime =\{n^\prime, P_y\}$
while $X_\alpha = \int_0^{T_\alpha}x(t)dt$ and $x(t)$ is the
x-coordinate of the electron which is defined by the classical
Hamilton equation Eq.(\ref{HamiltEquat}).

Expanding the integrand  in the vicinities of degeneration points
$P^{(n)}_k$ (where the resonant tunnelling takes place) with the
use of Eq.(\ref{Constants}) one re-writes
 the current, Eq.(\ref{CurrentConstants}), as follows:
\begin{eqnarray}
J=\frac{L_y}{L_x}2 e^2 \nu_0 V_0\sum_{n,k}
\Big( \frac{c P^{(n)}_y(k) }{ e H T_1}\Big)\nonumber \\
\times \int_{-\infty}^\infty d P_y \frac{|t|^2 }{(( v_{-} P_y/2
\hbar \omega_{12})^2 +|t|^2 }\nonumber \\ \times \frac{f_0(E_n+v_1
P_y))-f_0(E_n+v_2 P_y))}{(v_{2}-v_{1})P_y} \label{current2}
\end{eqnarray}
Here the summation is over all the degeneration points;  $\tau
=\nu^{-1}$ is the electron-impurity relaxation time,  $\omega_{12}
=1/ \sqrt{T_1 T_2}$ and
 $v_{1,2}$ are
 the  velocities  of the left (1) and right (2) edge states at $|t|=0$
while $T_{1}(E,P_y)$ and $T_{1}(E,P_y)$ are the times of  electron
motion
 between points $y_a$ and $y_b$ along the left and right classical orbits
 shown in Fig.\ref{orbits}, respectively:
\begin{eqnarray}
v_{1,2} &=& \frac{d E_{1,2}^{(0)}}{d P_y} =
 -\frac{\partial \theta_{1,2}/\partial P_y}{\partial \theta_{1,2}/\partial
 E};\nonumber \\
 T_{1,2}&= & \hbar \frac{\partial \theta_{1,2}}{\partial
 E}=\frac{1}{\omega_H}\left(\frac{\pi}{2}\pm \arcsin \frac{P_y}{p_E}\right)
\label{velocity T}
\end{eqnarray}
All the above-mentioned quantities are taken at $E=E_n,
 \;\;P_y=P_y^{(n)}(k)$.

 The first resonant  term of the integrand is due to the resonant
 transmission of  electrons between left and right edge states
 skipping along the lateral junction:  at the degenerate points
 $P_y=P_k^{(n)}$ the widths of the left and right wells (which are
 created by the magnetic field) are of such  values that
 the electron energies in them (at $|t|=0$)
 coincide  causing resonant transmissions between the wells (see Eq.\ref{Constants}
 and the text below it).

For the case that the temperature satisfies the inequality $kT
\gtrsim |t|\hbar \omega_H$ one may expand the Fermi distribution
functions with respect to $v_{1,2}P_y$ and obtain the current as
follows:
\begin{eqnarray}
&J&=|t|\frac{L_y  }{L_x}\frac{\sigma_{0}V_0}{ (\omega_H
\tau)^2}\nonumber \\ &\times& \frac{ \hbar\omega_H }{4 \pi^2
T}\sum_{n} \cosh^{-2}\left[\frac{\hbar \omega_H(n+1/2)
-\varepsilon_F}{2 T}\right] \label{currentfinal}
\end{eqnarray}
where $\sigma_0 n_F e^2 \tau/m$ is the Drude conductivity , $n_F
=p_F^2/\hbar^2$ is the electron density. While writing this
equation the summation with respect to $k$ was changed to
integration according to Eq.(\ref{variablechange}).

\begin{figure}
  \centerline{\includegraphics[width=8.0cm]{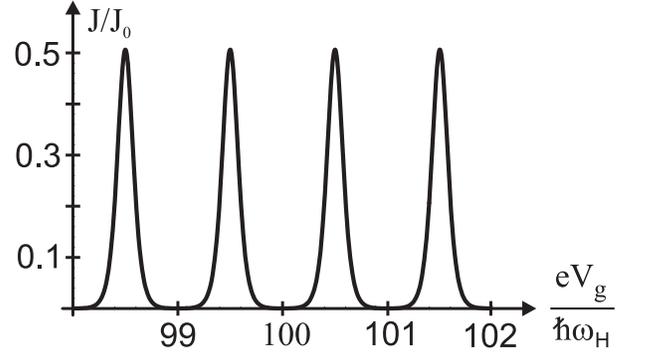}}
  \caption{Dependence of the current (which flows perpendicular to the lateral junction
  under magnetic field) on  the gate voltage $V_g$; the current is
  normalized to  $J_0=|t|\big(\sigma_0/\omega_H \tau)^2\big) V_0
  L_y/L_x$.
  The numerical
  calculations are carried out for  $ T/\hbar \omega_H =0.05$ and
  the semiclassical parameter
  $\hbar \omega_H/\varepsilon_F =10^{-2}$}
\label{current}
  \end{figure}

As one sees from Eq.(\ref{currentfinal}), at $T \ll \hbar \omega$
the current oscillates with a giant amplitude under a change of
the magnetic field or the gate voltage  (see Fig.\ref{current}).

If $T \gg \hbar \omega$ the summation with respect to $n$ may be
changed to integration,  $\sum_n ... \rightarrow \int dn ...$, and
the  current oscillations are smoothed out and the current becomes
of the conventional form.

%
%

In conclusion, it is shown that kinetic properties of a dirty 2DEG
system with lateral junction under magnetic field is extremely
sensitive to actions of  external fields. In particular, even a
rather weak  variation of the magnetic filed or the  gate voltage
causes giant oscillations of the charge current flowing
perpendicular to the junction. The theoretical analysis based on
quantum resonance tunnelling of quasi-particles skipping along the
junction is in a good agreement with experimental data: the period
of the conductance oscillations,  the position and the value of
the conductance maxima correspond to the observations presented in
Rwf.\ref{kang}.

\appendix
\section{Wave functions and dispersion equation for
quasi-particles skipping along lateral junction.
\label{Appendmatching}} Quantum dynamics of electrons  with a
lateral junction under magnetic field is described by the wave
function $\Psi(x,y)$ satisfying the two-dimensional Schr\"odinger
equation:
%
%
\begin{eqnarray}
 &-&\frac{\hbar^2}{2 m} \frac{\partial^2 \Psi}{\partial x^2}+
 \Big[\frac{1}{2 m}\big(-i \hbar \frac{\partial }{\partial y}+
 \frac{e H x}{c}\big)^2 \nonumber \\
 &+&\frac{m \omega_1^2}{2}y^2 + V (x)-E\Big]\Psi =0,
\label{Shcroedinger}
\end{eqnarray}
%
%
where the  gauge is used for which the vector-potential  ${\bf A}
= (0,Hx,0)$,

 The semiclassical solutions of the
above equation on the left and right sides of   the junction
($x_l<x<0$ and $0 <x<x_r$, respectively) are

\begin{eqnarray}
\Psi_1=\frac{C_1(n,P_y)}{\sqrt{p(x)/m}} \Big[
\exp\left\{\frac{i}{\hbar}\int_{x_1}^x
p(x^\prime)dx^\prime)-\frac{\pi}{4}\right\} + h.c.\Big]\nonumber \\
\Psi_2=\frac{C_2(n,P_y)}{\sqrt{p(x)/m}} \Big[
\exp\left\{\frac{i}{\hbar}\int_{x}^{x_2}
p(x^\prime)dx^\prime)-\frac{\pi}{4}\right\} + h.c.\Big]
\nonumber\\
 \label{wavefunctions}
\end{eqnarray}
%
where quantum numbers $n$ and $P_y$ are the band number and the
conserving projection momentum, respectively  while
%
%
\begin{eqnarray}
x_{1,2}=\frac{c}{eH}\left(\mp \sqrt{2 m E}-P_y\right);\;\;
  \nonumber \\
 p(x)=\sqrt{2 m E -\left(P_y+\frac{eH}{c}x\right)^2}
\label{momentaandphi}
\end{eqnarray}
%
%
Here $x_{1,2}$ are the turning points. The dependence of the
constant factors $C_1$, $C_2$ and  the quasi-particle dispersion
law $E_n=E_n(P_y)$ are found   by matching the above wave
functions at the lateral barrier and their  normalization  as it
is shown below.

In the vicinity of the junction $|x|\ll R_H$,  one may expand the phases
 of the wave functions Eq.(\ref{wavefunctions}) in $|x|$ and see that they are
 incoming and outgoing plane waves $\exp\{(\pm i px/\hbar)\}$ the constant factors at which
 $A_{1,2}$ and $B_{1,2}$ are
%
%
\begin{eqnarray}
A_1=  C_1  \exp\{i \big(\theta_1-\frac{\pi}{4}\big)\};\;
B_1=  C_1  \exp\{-i \big(\theta_1-\frac{\pi}{4}\big)\};\nonumber \\
A_2=  C_2  \exp\{i \big(\theta_2-\frac{\pi}{4}\big)\};\; B_2=C_2
\exp\{-i \big(\theta_2-\frac{\pi}{4}\big)\};\nonumber \\
\label{plainwavevactors}
\end{eqnarray}
%
where
%
%
\begin{eqnarray}
\theta_1=\int_{x_1}^0 p(x^\prime)dx^\prime; \hspace{0.2cm}
\theta_2=\int^{x_2}_0 p(x^\prime)dx^\prime; \label{thetas}
\end{eqnarray}
%
Changing variables in the integrals here one finds
Eq.(\ref{theta}) of the main text.

The found plane waves  undergo    two-channel scattering  at  the
junction  and    the  constant  factors  at the outgoing functions
are coupled with the incoming ones
 with  a $2\times 2$ scattering  unitary  matrix
which  is written  in  the  general case as follows:
\begin{equation}
\left(\begin{matrix} B_1\\
B_2
\end{matrix} \right)=e^{i\Phi}\left(\begin{matrix} r& t\\
-t^{\ast} & r^{\ast}
\end{matrix} \right)\left(\begin{matrix} A_1\\
A_2
\end{matrix} \right),
\label{taumatrix}
\end{equation}
where $t$ and $r$ are the probability amplitudes for the incoming
electron to pass through and to be scattered back at the junction,
respectively, $|t|^2+|r|^2=1$.
%
%
%

Using Eqs.(\ref{plainwavevactors},\ref{taumatrix}) one finds the
set of equations that couples the constant factors in wave
functions Eq.(\ref{wavefunctions}):
\begin{eqnarray}
\left(e^{-i(\bar{\theta}_1+\Phi)}- re^{i\bar{\theta}_1} \right)C_1
-te^{+i\theta_2}C_2=0; \nonumber \\
t^{\ast}e^{i\bar{\theta}_1}C_1 +\left(e^{-i(\bar{\theta}_2+\Phi)}-
r^{\ast}e^{i\bar{\theta}_2} \right)C_2=0;
\label{matchequationfinal}
\end{eqnarray}
where $\bar{\theta}_{1,2}=\theta_{1,2}-\pi/4$. Equating the
determinant of equation Eq.{\ref{matchequationfinal}} to zero and
using the inequality $|t|\ll 1$ one
 finds  dispersion equation
Eq.(\ref{dispersioneq}) of the main text.

Using equations Eq.(\ref{dispersioneq}) one easily finds the
quasi-particle  dispersion law in the vicinity of  points of
degeneration, $P_y=P_k^{n}$, as follows:
\begin{eqnarray}
\delta E_{\pm}=\frac{1}{2} \Big(\delta P_y v_{+} \pm \sqrt{(\delta
P_y v_{-})^2 +4(|t| \hbar \omega_{1,2})^2} \Big)
 \label{liftdegeneracy}
\end{eqnarray}
where $v_{\pm} = v_2-v_1$ and $\omega_{1,2}=1/\sqrt{T_1 T_2}$
while definitions of the velocities $v_{1,2}$ and times $T_{1,2}$
at  degeneration points are given by Eq.(\ref{velocity T}).

Normalization of  the wave function Eq.(\ref{wavefunctions})
 to unity gives the second independent equation for constants $C_{1,2}$:
\begin{eqnarray}
T_1 |C_1|^2+T_2 |C_2|^2=1
 \label{normalization}
\end{eqnarray}
where $T_{1,2}(E,P_y) =\hbar\; \partial \theta_{1,2}/\partial E$ are times of electron motion
along  classical orbits 1 and 2.

For the case under considerations $|t|^2\ll 1$, using
Eqs.(\ref{matchequationfinal},\ref{normalization}) one finally
finds
\begin{eqnarray}
|C_1(E_n,P_y)|^2 = \frac{|t|^2}{4 T_2 \cos^2\bar{\theta}_1 +
|t|^2 T_1}\nonumber \\
|C_2(E_n,P_y)|^2 = \frac{ 4 \cos^2\bar{\theta}_1 }{4 T_2
\cos^2\bar{\theta}_1 + |t|^2T_1}
 \label{Constants}
\end{eqnarray}
where functions $\bar{\theta}_{1,2}(E_n(P_y),P_y) $ are determined
by Eq.(\ref{theta}).

Using Eq.(\ref{liftdegeneracy}) one easily finds
$\cos^2\bar{\theta}_1 \approx |t|^2T_1/4 T_2$  at degeneracy point
$P_y= P_k^{(n)}$ and hence $|C_{1,2}(E_n,P_y)|^2=1/2 T_{1,2}$ that
is a resonant tunnelling takes place at these points.

\section{Matrix elements of  applied voltage and quasi-particle velocity. \label{matrixelements}}

1. Matrix elements of the applied voltage Eq.(\ref{voltage}) are
\begin{eqnarray}
V_{n,n^\prime}(P_y)=\int_{x_1}^{0} \Psi^\ast_{1,
\kappa^\prime}(x)\Psi_{1,\kappa}(x)dx
 \label{Vmatr}
\end{eqnarray}
where $\kappa =(n, P_y)$ and $\kappa^\prime =(n^\prime, P_y)$.
Using Eq.(\ref{wavefunctions})  one finds
\begin{eqnarray}
V_{n,n^\prime}(P_y)= V_0
C_1^\ast(\kappa) C_1(\kappa^\prime)\hspace{1cm}\nonumber\\
\times\frac{\sin\{T_1[E_n(P_y)-E_{n^\prime}(P_y)]/\hbar)\}}{E_n(P_y)-E_{n^\prime}(P_y)},
\;\; n \neq n^\prime \label{Vnn}
\end{eqnarray}

In the vicinity of degeneration points one has
$|E_n(P_y)-E_{n^\prime}(P_y)|\ll \hbar \omega_H$ and hence
Eq.(\ref{Vmatr}) may be written as follows:
\begin{eqnarray}
V_{n,n^\prime}(P_y)=\hbar V_0 T_1 |C_1(n,P_y)|^2
\label{VnnDegeneration}
\end{eqnarray}

 2. Matrix elements of the quasi-particle velocity are
\begin{eqnarray}
v^{(x)}_{n,n\prime}&=&\Big\{\int_{x_1}^0
\Psi_{1,\kappa}^\ast\frac{i}{\hbar}\big[\hat{H},x\big]\Psi_{1,\kappa^\prime}dx
\nonumber
\\
 &+&
\int^{x_2}_0
\Psi_{2,\kappa}^\ast\frac{i}{\hbar}\big[\hat{H},x\big]\Psi_{2,\kappa^\prime}dx\Big\}
 \label{Matrixvelocity}
\end{eqnarray}

Using  the explicit form of the semiclassical wave functions
Eq.(\ref{wavefunctions}) one finds the  velocity matrix elements
as follows
\begin{eqnarray}
v^{(x)}_{n,n\prime}=i
\frac{E_n(P_y)-E_{n^\prime}(P_y)}{\hbar}\sum_{\alpha=1}^2
C_\alpha^\ast(\kappa)C_\alpha(\kappa^\prime)
\hspace{1cm} \nonumber \\
\times \int_0^{T_\alpha}x^{(\alpha)}(t)\exp\Big\{i
\frac{\big[E_{n^\prime}(P_y)-E_{n}(P_y)\big](T_\alpha-t)}{\hbar}\Big\}d
t \nonumber \\
 \label{matrixvelocityfinal}
\end{eqnarray}
where $x(t)$ is defined by the classical Hamilton equation:
\begin{eqnarray}
\frac{d \bf{p}}{d t}=\frac{e}{c}\big[\bf{v} \bf{H}\big]
 \label{HamiltEquat}
\end{eqnarray}
while $x^{(1)}(t)$ and $x^{(2)}(t)$ are coordinates of
semiclassical packets moving along the left and right sections of
the closed orbit, respectively (see Fig.\ref{orbits}). They are
defined in such a way that  their motion starts at the beginning
of the corresponding section: e.g., for the motion along the
closed orbit in Fig.\ref{orbits} $x^{(1)}(0)=0$, $y^{(1)}(0)=y_b$
and $x^{(2)}(0)=0$, $y^{(2)}(0)=y_a$.

Using  inequality $|E_n(P_y)-E_{n^\prime}(P_y)|\ll \hbar
\omega_H$, near degeneration points  one may write
Eq.(\ref{matrixvelocityfinal}) in the form:
\begin{eqnarray}
v_{n,n^\prime}= i
\frac{E_n(P_y)-E_{n^\prime}(P_y)}{\hbar}\nonumber
\\
\times \sum_{\alpha=1}^2
C_\alpha^\ast(\kappa)C_\alpha(\kappa^\prime)\int_0^{T_\alpha}x^{(\alpha)}(t)dt
 \label{mevelocitydegeneration}
\end{eqnarray}


\begin{thebibliography}{99}
\bibitem{Berezinskii} V. L. Berezinskii, Sov. Phys. JETP, {\bf 32}, 493
(1971).
\bibitem{KT} J. M. Kosterlitz and D. J. Thouless,
 Journal of Physics C: Solid State Physics, {\bf 6}, 1181 (1973).
\bibitem{Heinzel} Th. Heinzel,  Mesoscopic Electronics in Solid
State Nanostructures, Wiley-VCH (2003).
\bibitem{Dittrich} W. Zwerger, \emph{Theory of Coherent Transport} in
Th. Dittrich, G-L. Ingold, G. Sch\"{o}n, P. H\"{a}nggi, B. Kramer,
and W. Zwerger,  \emph{Quantum Transport and Dissipation},
Wiley-VCH (1998).
\bibitem{QH} B. Kramer, 1998 \emph{Quantization of Transport},  in Th. Dittrich,
G-L. Ingold, G. Sch\"{o}n, P. H\"{a}nggi, B. Kramer, and W.
Zwerger, \emph{Quantum Transport and Dissipation}, Wiley-VCH
(1998).
\bibitem{Devoret} M. H. Devoret , J. M. Martinis, and J. Clarke, 1985 Phys. Rev.
Lett. 55 (1908); J. M. Martinis, M. H. Devoret, and J. Clarke
 Phys. Rev. B 35, 4682 (1987).
\bibitem{Beenakker} C. W. J. Beenakker and H. van Houten, Solid State
Physics, 441 (1991).
\bibitem{kang} W. Kang, H.L. Stormer, L.N. Pfeifer, K.W. Baldwin, and K.W.
West, Letters to Nature, {\bf 403}, 59 (2000).
\bibitem{barrier} A.M. Kadigrobov, M.V. Fistul, and K.B. Efetov,  Phys. Rev. B
{\bf 73}, 235313 (2006).
\bibitem{graphene} A.M. Kadigrobov,
arXiv:1609.06648, to be published in Low Temperature Physics {\bf
43}, No 1 (2017).
\bibitem{Cohen} Morrel H. Cohen and L.M. Falikov, Pfys. Rev. Lett.
{\bf 6}, 231 (231).
\bibitem{KaganovSlutskin} M.I. Kaganov and A.A. Slutskin, Physics
Reports {\bf 98}, 189 (1983).
\bibitem{FTT} A.A. Slutskin and A.M. Kadigrobov, Soviet Physics -
Solid State, {\bf 9}, 138 (1967).
\bibitem{Slutskin} A.A. Slutskin,  Sov. Phys. JETP Lett. {\bf 26}, 474 (1968).
\bibitem{reflection} A.A. Slutskin and A.M. Kadigrobov, JETP Lett. {\bf 32}, 338 (1980)
\bibitem{physica} A.A. Slutskin and A.M. Kadigrobov, Physica B \& C {\bf
108}, 877 (1981).
 \bibitem{Peschanski} Y.A. Kolesnichenko, V.G. Peschanski,   Fizika Nizkikh Temperatur {\bf 10}, 1141
 (1984)
\bibitem{Koshkin}  A.M. Kadigrobov and I.V. Koshkin, Sov. J. Low
Temp. Phys. {\bf 12}, 249 (1986).
\bibitem{barrierperp} A.M. Kadigrobov and  M.V. Fistul, J. Phys.:
Condens. Matter {\bf 28}, 255301 (2016).



\end{thebibliography}
 \end{document}